\documentstyle[12pt,psfig]{article}
\catcode`\@=11
\long\def\@makefntext#1{
\protect\noindent \hbox to 3.2pt {\hskip-.9pt
$^{{\ninerm\@thefnmark}}$\hfil}#1\hfill}                

\def\@makefnmark{\hbox to 0pt{$^{\@thefnmark}$\hss}}  

\def\ps@myheadings{\let\@mkboth\@gobbletwo
\def\@oddhead{\hbox{}
\rightmark\hfil\ninerm\thepage}
\def\@oddfoot{}\def\@evenhead{\ninerm\thepage\hfil
\leftmark\hbox{}}\def\@evenfoot{}
\def\sectionmark##1{}\def\subsectionmark##1{}}

\renewcommand{\thefootnote}{\fnsymbol{footnote}}

\newcounter{sectionc}\newcounter{subsectionc}\newcounter{subsubsectionc}
\renewcommand{\section}[1] {\vspace*{0.6cm}\addtocounter{sectionc}{1}
\setcounter{subsectionc}{0}\setcounter{subsubsectionc}{0}\noindent
        {\normalsize\bf\thesectionc. #1}\par\vspace*{0.4cm}}
\renewcommand{\subsection}[1] {\vspace*{0.6cm}\addtocounter{subsectionc}{1}
        \setcounter{subsubsectionc}{0}\noindent
        {\normalsize\it\thesectionc.\thesubsectionc. #1}\par\vspace*{0.4cm}}
\renewcommand{\subsubsection}[1]
{\vspace*{0.6cm}\addtocounter{subsubsectionc}{1}
        \noindent
{\normalsize\rm\thesectionc.\thesubsectionc.\thesubsubsectionc.
        #1}\par\vspace*{0.4cm}}

\newcounter{appendixc}
\newcounter{subappendixc}[appendixc]
\newcounter{subsubappendixc}[subappendixc]

\renewcommand{\appendix}[1] {\vspace*{0.6cm}
        \refstepcounter{appendixc}
        \setcounter{figure}{0}
        \setcounter{table}{0}
        \setcounter{equation}{0}
        \renewcommand{\thefigure}{\Alph{appendixc}.\arabic{figure}}
        \renewcommand{\thetable}{\Alph{appendixc}.\arabic{table}}
        \renewcommand{\theappendixc}{\Alph{appendixc}}
        \renewcommand{\theequation}{\Alph{appendixc}.\arabic{equation}}
        \noindent{\bf Appendix \theappendixc #1}\par\vspace*{0.4cm}}

\def\abstracts#1{{

\centering{\begin{minipage}{12.2truecm}\footnotesize\baselineskip=12pt\noindent
        \centerline{\footnotesize ABSTRACT}\vspace*{0.3cm}
        \parindent=0pt #1
        \end{minipage}}\par}}

\renewenvironment{thebibliography}[1]
        {\begin{list}{\arabic{enumi}.}
        {\usecounter{enumi}\setlength{\parsep}{0pt}
\setlength{\leftmargin 1.25cm}{\rightmargin 0pt}
         \setlength{\itemsep}{0pt} \settowidth
        {\labelwidth}{#1.}\sloppy}}{\end{list}}

\newcounter{itemlistc}
\newcounter{romanlistc}
\newcounter{alphlistc}
\newcounter{arabiclistc}

\newcommand{\fcaption}[1]{
        \refstepcounter{figure}
        \setbox\@tempboxa = \hbox{\footnotesize Fig.~\thefigure. #1}
        \ifdim \wd\@tempboxa > 6in
           {\begin{center}
        \parbox{6in}{\footnotesize\baselineskip=12pt Fig.~\thefigure. #1}
            \end{center}}
        \else
             {\begin{center}
             {\footnotesize Fig.~\thefigure. #1}
              \end{center}}
        \fi}

\newcommand{\tcaption}[1]{
        \refstepcounter{table}
        \setbox\@tempboxa = \hbox{\footnotesize Table~\thetable. #1}
        \ifdim \wd\@tempboxa > 6in
           {\begin{center}
        \parbox{6in}{\footnotesize\baselineskip=12pt Table~\thetable. #1}
            \end{center}}
        \else
             {\begin{center}
             {\footnotesize Table~\thetable. #1}
              \end{center}}
        \fi}

\def\@citex[#1]#2{\if@filesw\immediate\write\@auxout
        {\string\citation{#2}}\fi
\def\@citea{}\@cite{\@for\@citeb:=#2\do
        {\@citea\def\@citea{,}\@ifundefined
        {b@\@citeb}{{\bf ?}\@warning
        {Citation `\@citeb' on page \thepage \space undefined}}
        {\csname b@\@citeb\endcsname}}}{#1}}

\newif\if@cghi
\def\cite{\@cghitrue\@ifnextchar [{\@tempswatrue
        \@citex}{\@tempswafalse\@citex[]}}
\def\citelow{\@cghifalse\@ifnextchar [{\@tempswatrue
        \@citex}{\@tempswafalse\@citex[]}}
\def\@cite#1#2{{$\null^{#1}$\if@tempswa\typeout
        {IJCGA warning: optional citation argument
        ignored: `#2'} \fi}}

 1
 1
 1

\font\ninerm=cmr9

\def\Journal#1&#2&#3(#4){#1{\bf #2} (#4) #3.}

\def\NIMA{{\em Nucl. Inst. and Meth. }{\bf A}}

\def\PLB{{\em Phys. Lett. }{\bf B}}

\def\PRL{{\em Phys. Rev. Lett. }}
\def\PRD{{\em Phys. Rev. }{\bf D}}
\def\PRC{{\em Phys. Rep. }{\bf C}}
\def\ZPC{{\em Z. Phys. }{\bf C}}

\def\etal{{\it et al.}}

\begin{document}

\baselineskip=16pt
\centerline{\normalsize\bf A MEASUREMENT OF THE FORM FACTORS OF LIGHT}
\centerline{\normalsize\bf PSEUDOSCALAR MESONS AT A LARGE MOMENTUM TRANSFER}

\centerline{\footnotesize VLADIMIR SAVINOV\footnote{Contribution to the
PHOTON95 conference, Sheffield (1995)
\newline
e-mail: savinov@lns62.lns.cornell.edu}
}
\baselineskip=13pt
\centerline{\footnotesize\it School of Physics and Astronomy,
University of Minnesota}
\baselineskip=13pt
\centerline{\footnotesize\it Minneapolis, MN 55455, USA}
\centerline{\footnotesize (representing the CLEO collaboration)}

\vspace*{0.5cm}
\abstracts{
Using the CLEO-II detector,
we have studied the exclusive two-photon
production of the light pseudoscalar
mesons in a single tagged mode.
We report on a preliminary measurement of the $\pi^0$, $\eta$ and
$\eta^{\prime}$ electromagnetic form factors in the
$Q^2$ region from 2 ${\rm (GeV/c)}^2$ to 20 ${\rm (GeV/c)}^2$.}

\normalsize\baselineskip=15pt
\setcounter{footnote}{0}
\renewcommand{\thefootnote}{\alph{footnote}}

\section{Introduction}

The processes discussed in this report are
$e^+e^- \rightarrow e^+e^-P$
where $P$ stands for one of the light pseudoscalar mesons $\pi^0$, $\eta$ and
$\eta^{\prime}$.
To lowest order in perturbation theory this process can be described
as single meson production by means of the two space-like
photons emitted by an electron or a positron.
In our experiment we detect the decay products of the meson and either
the electron or positron which has scattered at an angle more than 20 degrees
from the $e^+e^-$ collision axis. From now on we refer to the detected
electron (positron) as the ``tag''.
The other scattered lepton remains undetected.
We measure production rate as a function of the squared momentum transferred by
the photon emitted by the tag.
The momentum transfer, $Q^2$, can be expressed in terms
of the tag parameters:
$Q^2 \equiv -q^2 = 2 E_{{\rm beam}} E_{{\rm tag}}
(1-{\rm cos}\theta_{{\rm tag}})$.
Here $E_{{\rm beam}}$ and $E_{{\rm tag}}$ are
energies of the tag before and after scattering
and $\theta_{{\rm tag}}$ is the scattering angle of the tag.
The study of the deviation of the production rate
from that which is predicted for point-like mesons gives a form
factor measurement.
This furnishes information about hadronic structure.
In our analysis we measure $\pi^0$, $\eta$ and $\eta^{\prime}$
form factors using
the following decay channels: $\pi^0 \rightarrow \gamma\gamma$,~
 $\eta \rightarrow \gamma\gamma$,~  $\eta \rightarrow \pi^+\pi^-\pi^0$,~
$\eta^{\prime} \rightarrow \rho^0\gamma$
and $\eta^{\prime} \rightarrow \pi^+\pi^-\eta
{}~(\eta \rightarrow \gamma\gamma)$.

\section{Apparatus and Data Sample}

The measurement described here utilizes data collected with the CLEO-II
general purpose magnetic detector\cite{CLEO-II:detector} which is
operated at the Cornell Electron Storage Ring
{}~(CESR). CESR is a symmetric $e^+e^-$ collider running at the center
of mass energy around 10.6 GeV.
The excellent performance of CESR has allowed us
to accumulate an $e^+e^-$ integrated luminosity of 2.88 ${\rm fb}^{-1}$.
In CLEO-II charged particles trajectories are measured with a set of three
concentric cylindrical drift chambers located inside a solenoidal magnetic
field.
They cover 95 $\%$ of the solid angle, and
the momentum resolution is
$\sigma_p/p (\%) = \sqrt{(0.15 p)^2 + (0.50)^2}$ ($p$ in GeV/c).
The electromagnetic calorimeter consists of 7800 CsI scintillation
crystals and is used for photon detection and electron identification.
The relative energy and angular resolutions in the
barrel part of the calorimeter are
$\sigma_E / E (\%) = 0.35/E^{0.75}+1.9 -0.1 E$ and
$\sigma_\phi({\rm mrad}) = 2.8/\sqrt{E}+2.5$ ({\rm E in GeV}), respectively.
The calorimeter covers
$98\%$ of the $4\pi$.
The time-of-flight (TOF) system (97 $\%$ of $4\pi$) consists of scintillation
counters
and is used primarily for trigger purposes.

CLEO-II has a three-level trigger system\cite{CLEO-II:trigger}
which consists of ten triggers suitable for different physics.
In our analysis we use data collected with two triggers.
The first trigger requires that both the tag and the decay products of the
meson are detected.
The signature of the tag includes a track reconstructed
by a fast track-finding processor, a hit in the TOF counter
and a shower of at least 500 MeV.
In order to identify the decay products of the meson
another shower of the energy above 500 MeV
or two separated showers (at least of 200 MeV of deposited energy each)
must be found in the
barrel part of the calorimeter.
The other trigger requires at least two tracks, one with
momentum transverse to the beam line ($p_\perp$) greater than
400 MeV/c and the other with $p_\perp$ greater than 200 MeV/c.
In addition, hits in two nonadjacent TOF counters and two separated
energy clusters (above 200 MeV each) must be found in the barrel part of the
detector.

\section{Event Selection}

Event selection criteria are optimized to identify
two-photon events where the only missing particle is a scattered electron
or positron. We require that the following conditions are satisfied:
\begin{itemize}
\item Both the charged track and the shower
(of the energy above 1 GeV) are found for the tag candidate.
The shower centroid and the projected track impact point agree to within
20 degrees as viewed from the interaction point.
\item Energy of the two-photon system is less than 60\%
of the Energy of the Center of Mass (ECM).
\item Hadronic system has no net charge.
\item Events with hadronic tracks have $E_{{\rm isolated}}$
less than 300 MeV where $E_{{\rm isolated}}$
is a total energy of the extra photon-like clusters in the calorimeter
not associated
with the process under consideration.
Hadrons interacting strongly in the material of the electromagnetic
calorimeter are primary sources of these clusters.
We require $E_{{\rm isolated}}$ to be equal to zero when no charged
tracks besides the tag candidate are found in the whole detector.
\item The scattering angle of the unobserved electron (positron) is less than
20 degrees. This angle is determined from the direction of the missing momentum
which we assume to be due solely to the undetected electron (positron).
The missing momentum ($\vec{P}_{{\rm missing}}$)
is a vector opposite to the vector sum of the momenta of
all reconstructed charged particles and photons.
\item $|\vec{P}_{{\rm missing}}|$ and  the missing energy ($E_{{\rm missing}}$)
which is calculated by
subtracting the total detected energy ($E_{{\rm visible}}$)
from the ECM of the experiment must satisfy the following criteria:
\begin{center}
$2.0 ~{\rm GeV} \leq E_{{\rm missing}} \leq 6.0 ~{\rm GeV}$

$2.0 ~{\rm GeV} \leq |\vec{P}_{{\rm missing}}| \leq 6.0 ~{\rm GeV}$

$-2.0 ~{\rm GeV} \leq (E_{{\rm missing}} - |\vec{P}_{{\rm missing}}|)
\leq 2.0 ~{\rm GeV}^*$
\end{center}
\end{itemize}

{\footnotesize
\noindent
$^*$ A tighter cut was imposed in the $\eta \rightarrow \pi^+\pi^-\pi^0$
analysis.  See section 4 for the details.}
\vspace*{0.3cm}

\begin{figure}[h]
\centerline{\psfig{figure=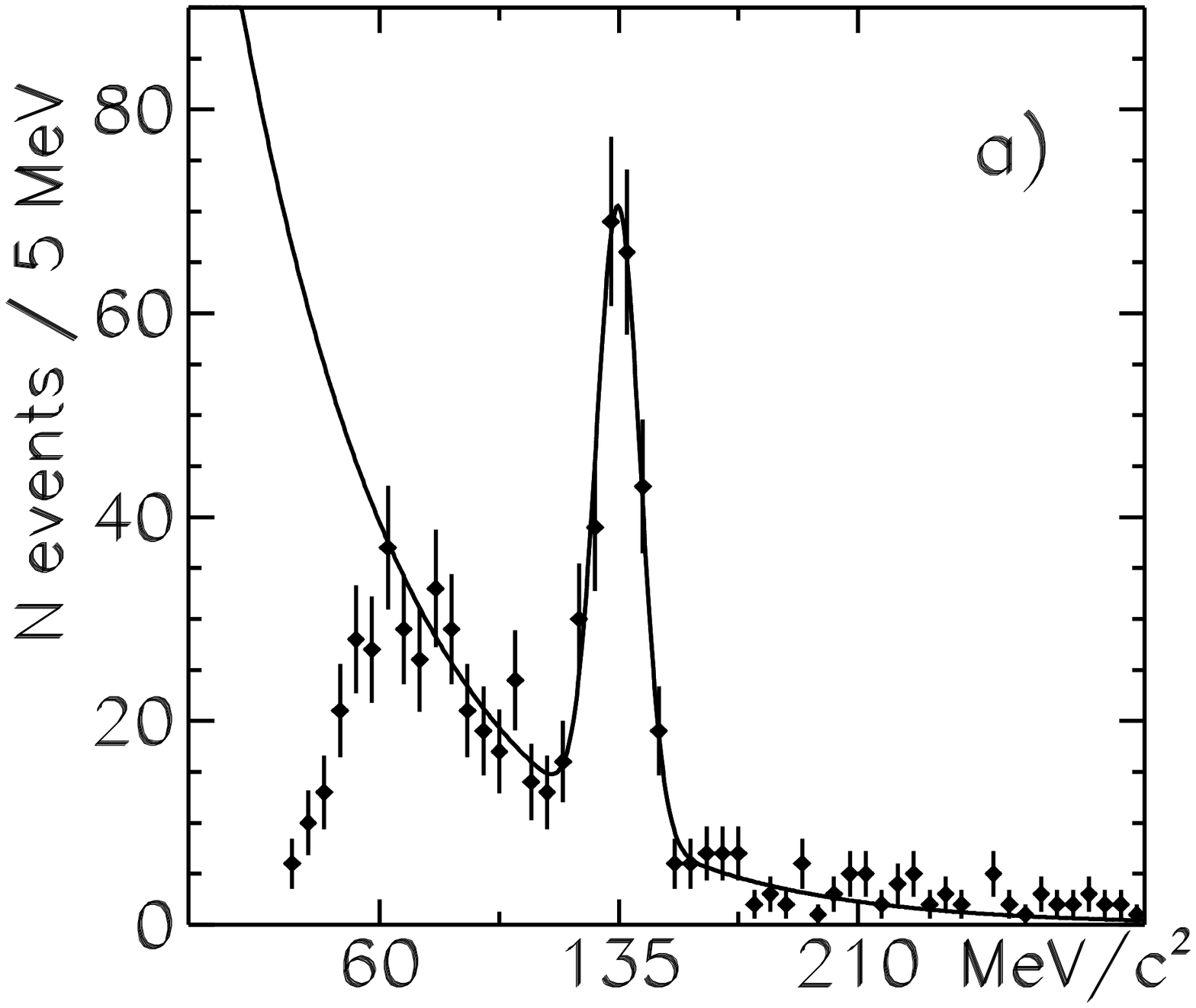,height=1.65in}
            \psfig{figure=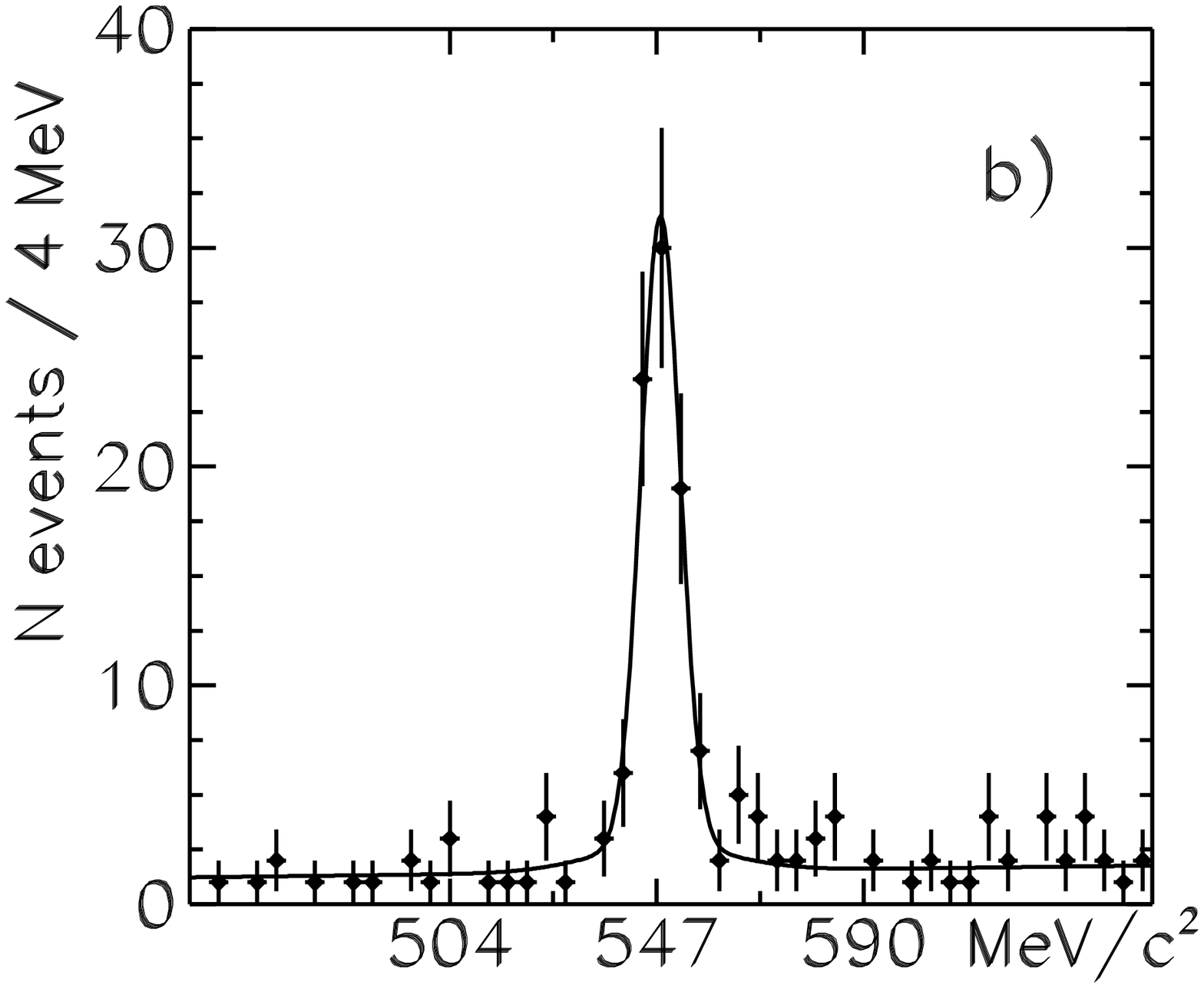,height=1.65in}
	    \psfig{figure=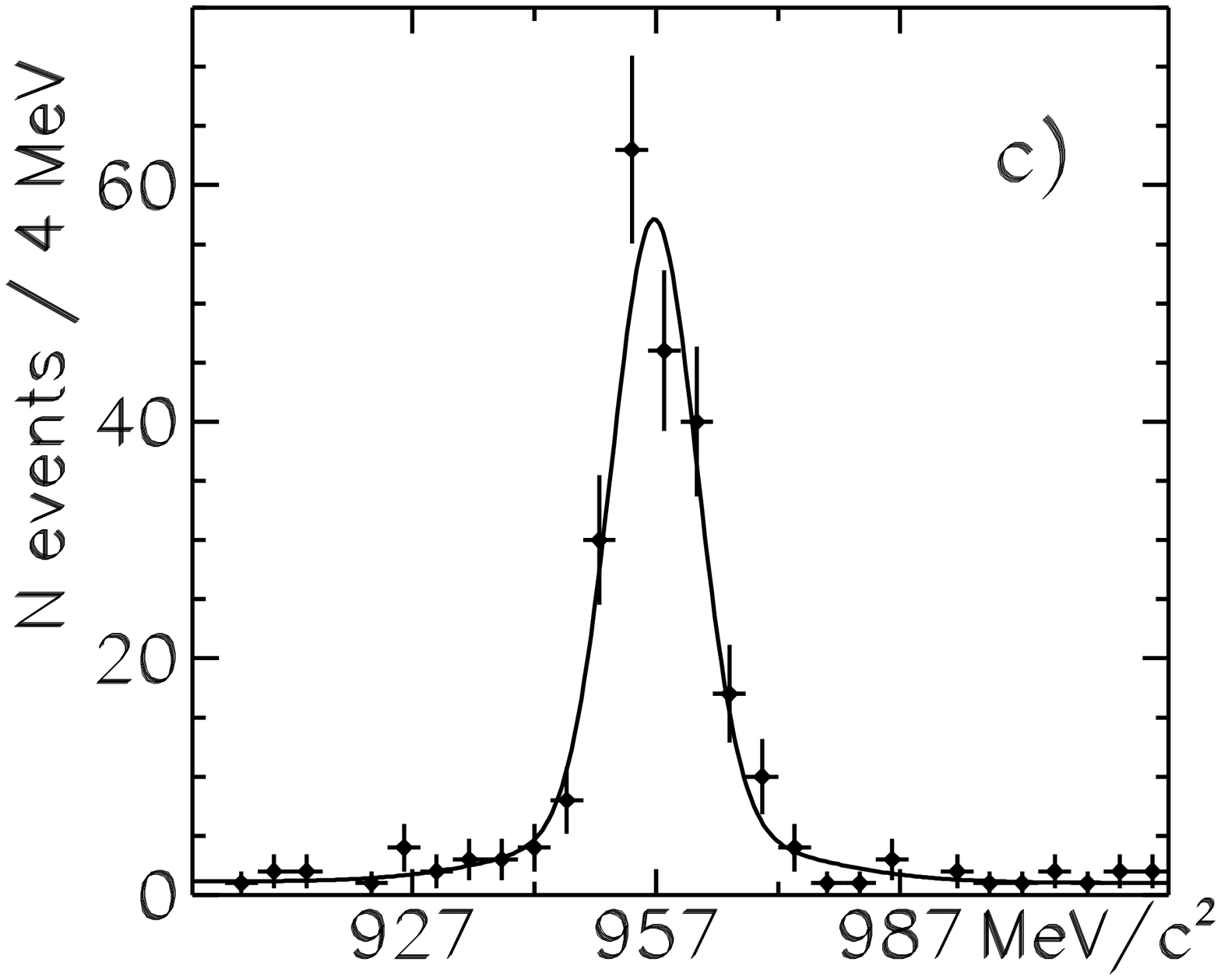,height=1.65in}}
\fcaption{The mass distributions for {a) $\pi^0 \rightarrow \gamma\gamma$},
{b) $\eta \rightarrow \pi^+\pi^-\pi^0$}
and {c) $\eta^{\prime} \rightarrow \eta\pi^+\pi^-
(\eta \rightarrow \gamma\gamma)$} signals.
The curves show the results of the fits described in the text.}
\label{fig:invmass}
\end{figure}

\noindent
We show the distributions of the hadronic system masses
for the candidate events in Fig.\ref{fig:invmass}.
The shapes of the signal are obtained using a GEANT-based detector simulation
program.
The shapes of the background are approximated by smooth functions that
are discussed below.

\section{Background Estimates}

Background events in which the final state
is not fully reconstructed may contribute to the
$\pi^0$, $\eta$ and $\eta^{\prime}$ signals.
In order to estimate this background
we have run an extensive Monte Carlo (MC)
simulation of different two-photon processes.
Typically our MC sample size exceeded our data by a factor of
30.
We identified the following process
which feeds into the $\eta \rightarrow \pi^+\pi^-\pi^0$ signal:
$\eta^{\prime} \rightarrow \eta\pi^0\pi^0$.
In order to supress this background a more restrictive cut
is applied in the $\eta \rightarrow \pi^+\pi^-\pi^0$ analysis:
$-0.6 ~{\rm GeV} \leq (E_{{\rm missing}} - |\vec{P}_{{\rm missing}}|)
\leq 0.6 ~{\rm GeV}$.

Simulation has been also performed for the single-photon annihilation process,
{\rm i.e.} $e^+e^- \rightarrow \gamma^* \rightarrow$ lepton pairs or hadrons.
We can make an independent estimate of the possible contribution from this
process by taking advantage of our knowledge of the tag charge.
In two-photon signal events the untagged electron (positron)
is scattered at a very small angle and is travelling along the
direction of the incident electron (positron) beam.
Thus, the direction of the missing momentum uniquely
determines the charge of the tag.
This is not the case for the single-photon annihilation events.
The observed rate for events with a wrong charge of the tag
is comparable to the charge misidentification probability of $\approx 0.5\%$.
This implies that the background from the annihilation processes is negligible.

\begin{figure}[h]
\centerline{
	\psfig{figure=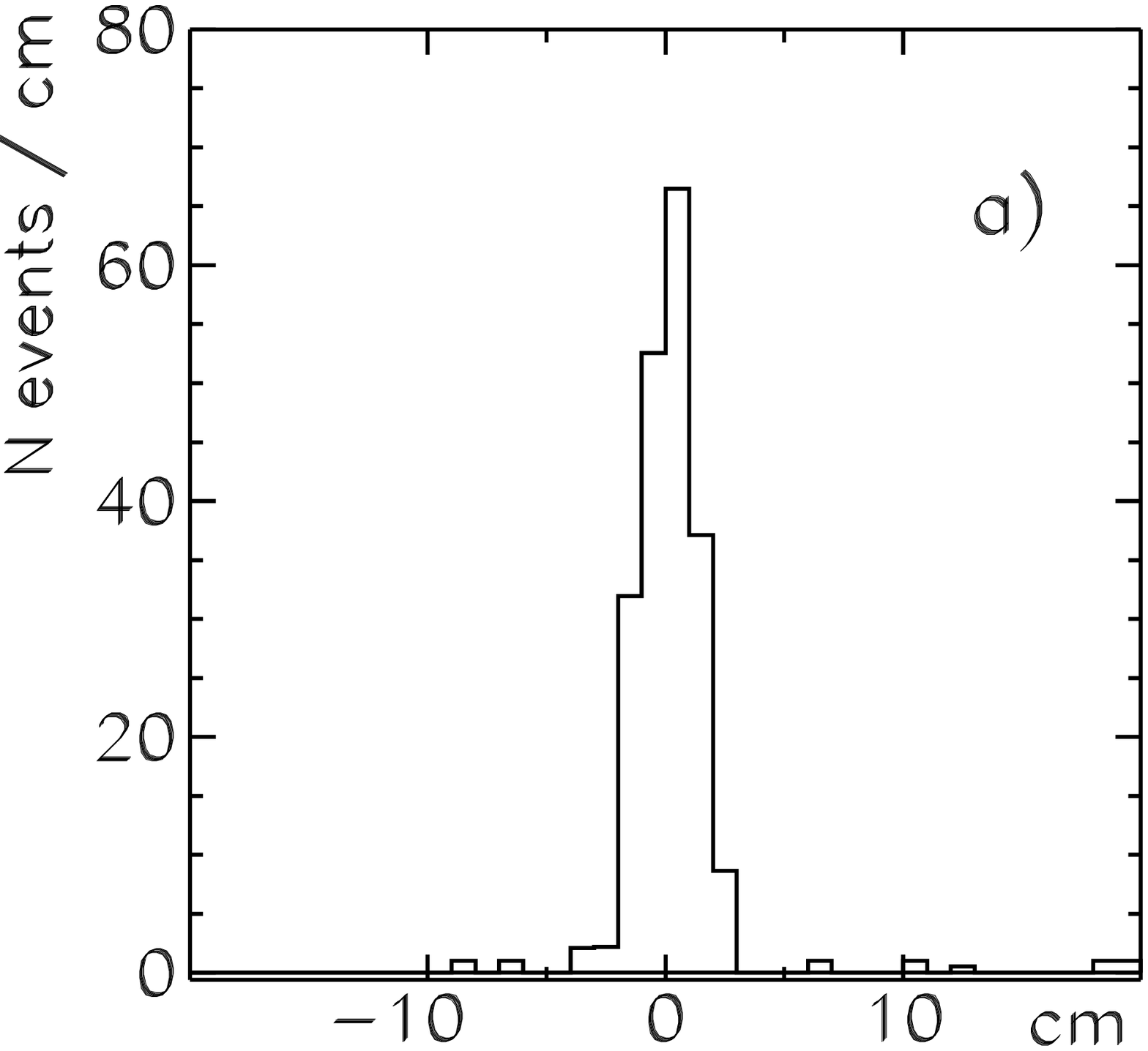,height=1.65in}
	\hspace{20mm}
	\psfig{figure=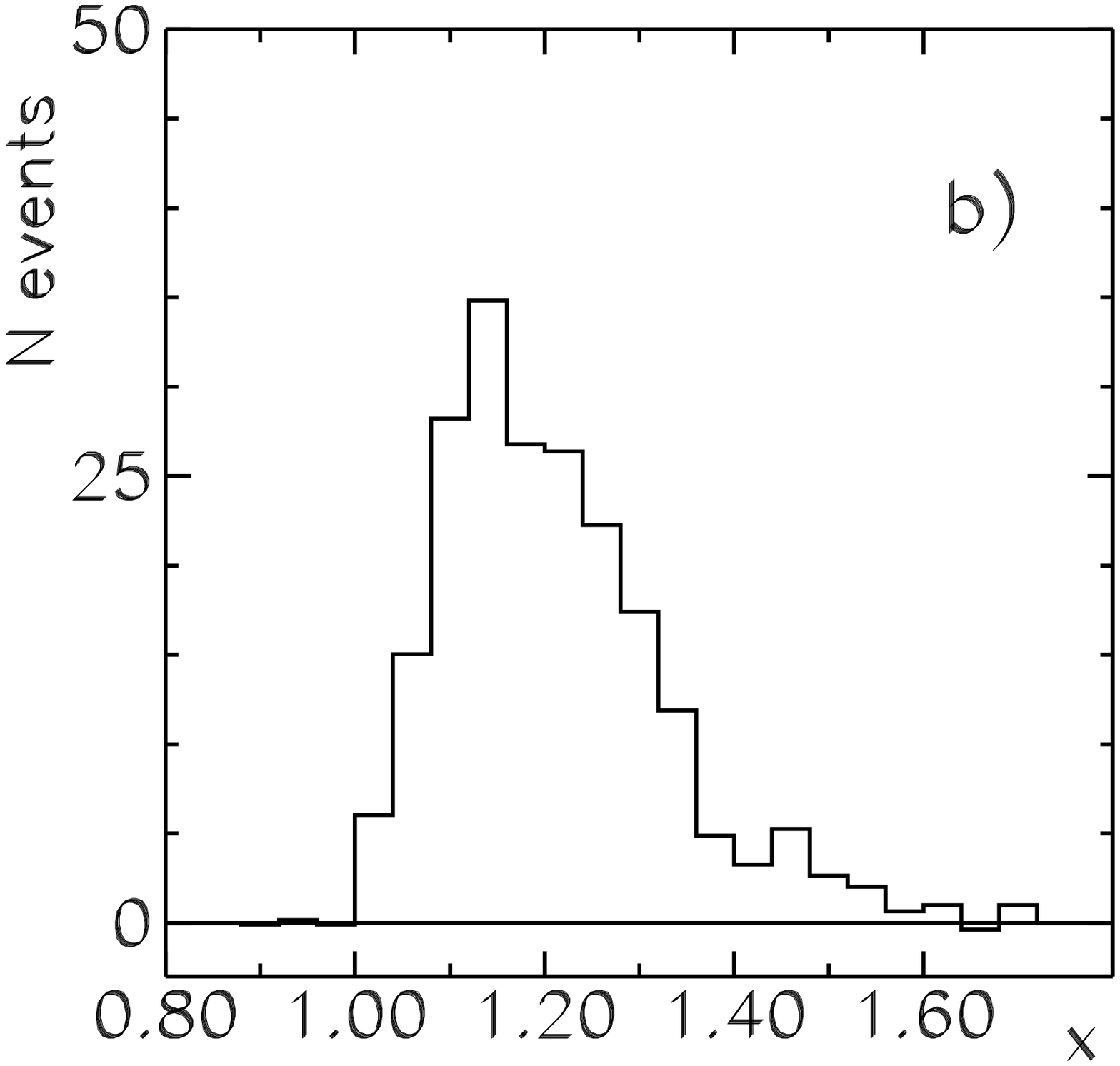,height=1.65in}}
\fcaption{a) The position of the event vertex along the beam line for
$\eta^{\prime}$ signal,
{}~b) $x = E_{{\rm visible}}/E_{{\rm beam}}$ for the $\pi^0$ signal
(background subtracted using sidebands in the mass distribution)}
\label{fig:visen}
\end{figure}

Electro-production of a meson in the field of a
beam-gas nucleus is also a possible source of background. We estimate
this background using distributions
of the event vertex position along the
beam line and $x = E_{{\rm visible}}/E_{{\rm beam}}$
which are shown in Fig.\ref{fig:visen}.
We assume that beam-gas events peak at {\it x} = 1.0 and are
uniformly distributed along the beam line.
{}From the number of events in the tails of the distributions
we conclude that beam-gas contamination is insignificant.

The total contribution from all background processes associated with resonance
production is estimated to be less than 1 event in each channel studied.

\noindent
The background contribution from all other processes
is approximated by the polinomial
of the first order except for the background under the $\pi^0$ signal peak.
This background (shown in Fig.1a)) comes from
the radiative Bhabha events, {\rm i.e.} $e^+e^- \rightarrow e^+e^-\gamma$.
In these events a $\gamma$ converts into $e^+e^-$ pair in the outer
layers of the drift chamber.
The 1.5 Tesla magnetic field causes a spatial separation of the produced
pair and the electromagnetic showers developed in the calorimeter
are indistinguishable from the showers caused by real photons when no
tracks are found.
We use an exponential to approximate contribution from this process.

\section{Analysis and Fits to the Form Factors}

\begin{figure}[h]
\centerline{\psfig{figure=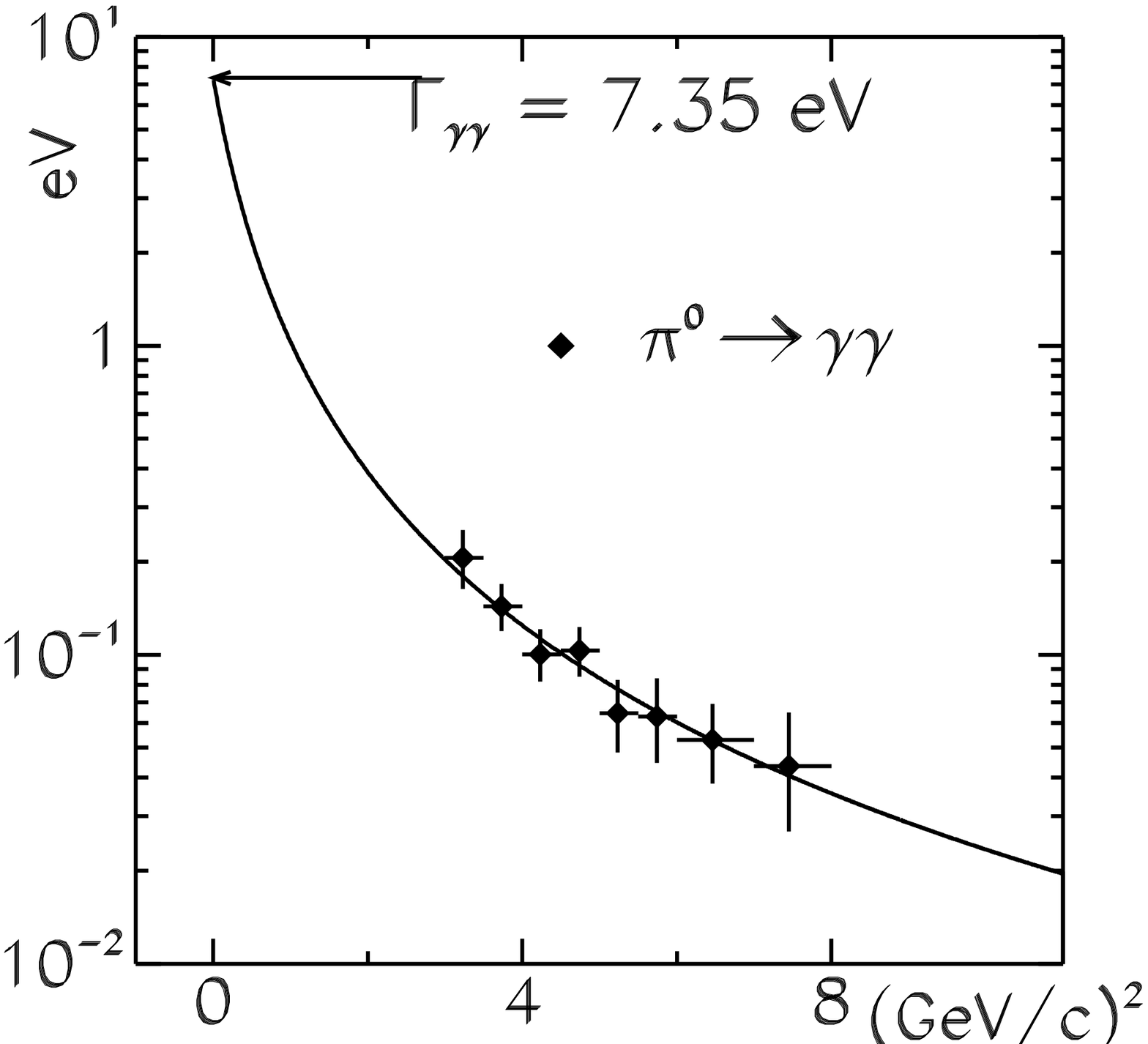,height=1.70in}
            \psfig{figure=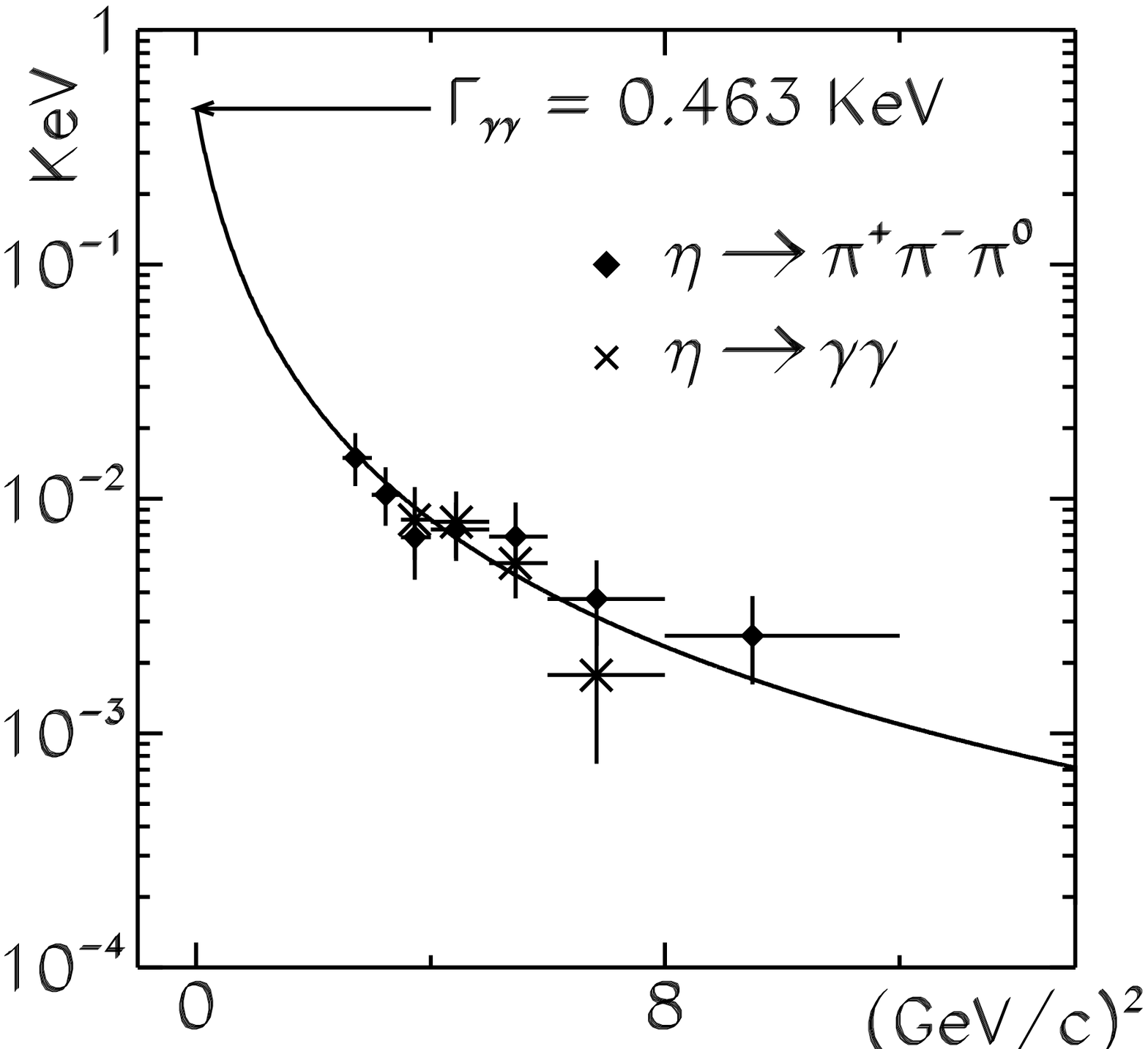,height=1.70in}
	    \psfig{figure=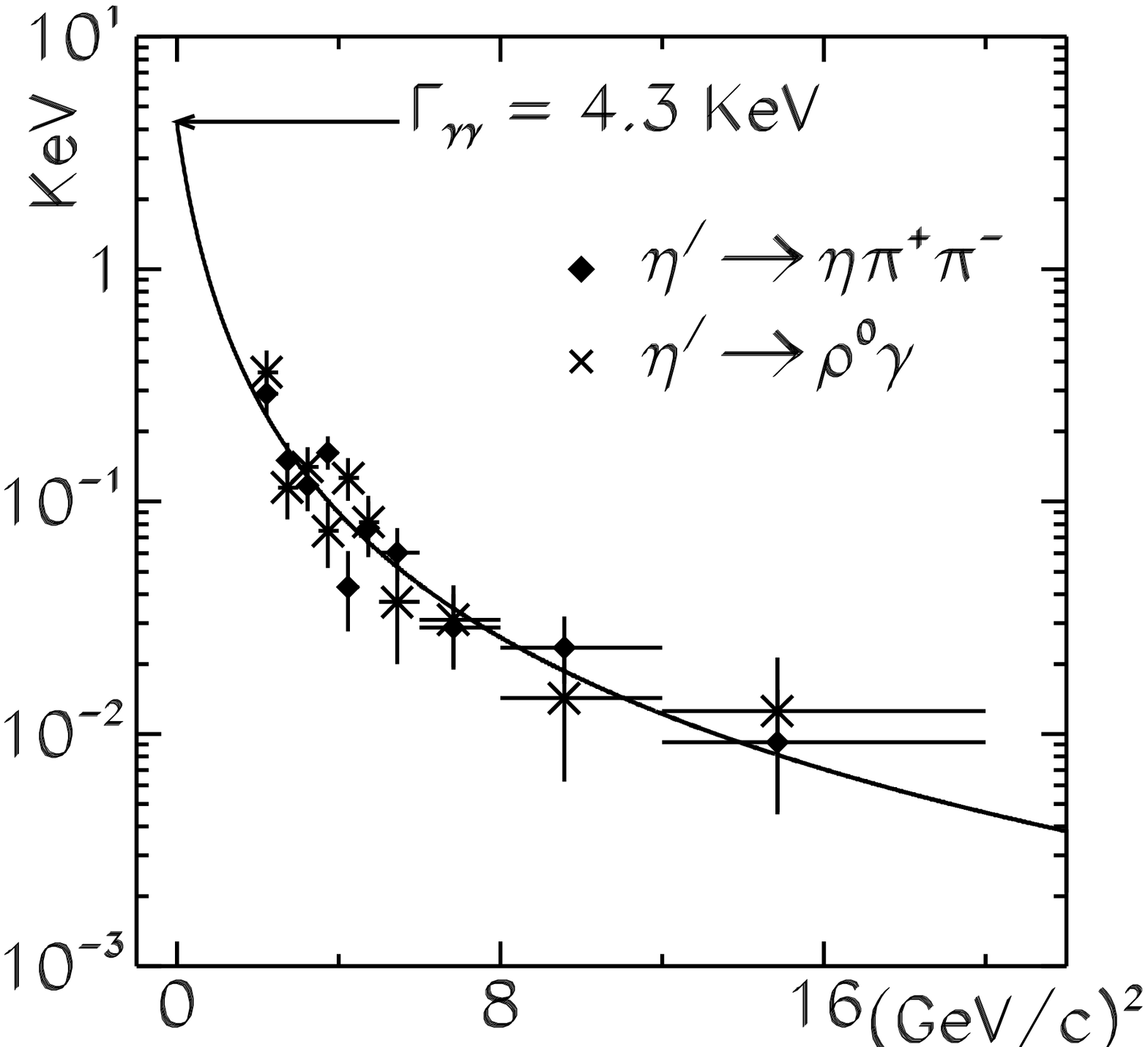,height=1.70in}}
\fcaption{The $\pi^0$,~$\eta$ and $\eta^{\prime}$ pseudoscalar
meson form factors (preliminary). Description of the pole mass fit can be
found in the text}
\label{fig:ff}
\end{figure}

In order to achieve the best possible resolution in $Q^2$, we employ
the following techniques:
\begin{itemize}
\item For
$\eta^{\prime} \rightarrow \pi^+\pi^-\eta  (\eta \rightarrow \gamma\gamma)$ and
$\eta \rightarrow \pi^+\pi^-\pi^0$ events a full kinematic fit is performed
where the photon pair is constrained to have exactly
$\eta$ or $\pi^0$ mass.
\item The tag scattering angle is measured from the centroid of the shower
associated with the tag
rather than from the reconstructed charged track.
\item The magnitude of the transverse momentum of the tag
($p^{{\rm tag}}_{\perp}$)
is constrained to be equal to the $p_{\perp}$ of the hadronic system.
\item The energy of the tag ($E_{{\rm tag}}$) is obtained using scattering
angle and $p^{{\rm tag}}_{\perp}$.
Since shower leakage at small angles is significant,
this method gives a better $E_{{\rm tag}}$ resolution than
the endcap calorimeter energy measurement.
\end{itemize}
For the purposes of obtaining the detection efficiency
we use MC generators based on the formalism of Budnev\cite{Budnev}
with a simple double pole vector meson dominance (VMD)
form factor incorporated.
We measure trigger efficiencies by using redundancies between different
triggers. Overall detection efficiency
is obtained in bins of $Q^2$ and this efficiency
changes from 5$\%$ to up to 30$\%$ as $Q^2$ grows.
We extract the $Q^2$ dependence of the form factor by comparing
the measured production rate
with that which is predicted for point-like pseudoscalar mesons.
Our preliminary results are shown in Fig.\ref{fig:ff}.
\noindent
The choice of the bin centers in $Q^2$ is made on the basis of the VMD MC
simulation. We interpret our results in terms of $\Lambda_p$, the mass
parameter which governs the $Q^2$ evolution of the form factor.
This parameter is obtained by fitting our data with a function of the
following form:
\begin{center}
${\cal F}^2(Q^2,0)*M^3/64\pi = \Gamma_{\gamma\gamma}/(1+Q^2/\Lambda_p^2)^2 $
\end{center}
Here ${\cal F}^2(Q^2,0)$ is the square of the form factor which is a
function of $Q^2$, $M$ is the mass of the meson and
$\Gamma_{\gamma\gamma}$ is the two-photon
width of the resonance (7.35 eV, 0.463 KeV and 4.3 KeV for
$\pi^0$, $\eta$ and $\eta^{\prime}$, respectively\cite{PDG:92}).
\begin{table}
\begin{center}
\tcaption{Summary of the results on pole mass $\Lambda_p$ [MeV],
governing $Q^2$ evolution of the form factors}
\smallskip
\smallskip
\label{tab:massresults}
\begin{tabular}{|c|c|c|c|c|}
\hline
& ~~~~Lepton-G\cite{LEPTONG:ff}~~~~ & ~~~~~~TPC/2$\gamma$\cite{TPC:ff}~~~~~
& ~~~~~CELLO\cite{CELLO:ff}~~~~~~ & CLEO (this study) \\
\hline
$\pi^0$ 	  &	         & 	        & 748 $\pm$ 30 & 771 $\pm$
18 \\
\hline
$\eta$  	  & 720 $\pm$ 90 & 700 $\pm$ 80 & 839 $\pm$ 63 & 787 $\pm$
23 \\
\hline
$\eta^{\prime}$ &                & 850 $\pm$ 70 & 794 $\pm$ 44 & 821 $\pm$
15 \\
\hline
\end{tabular}
\end{center}
\end{table}
We estimate the average squared momentum of the second photon to be
less than $0.001 $~${\rm (GeV/c)}^2$.
It is predicted\cite{BL:81} that in the asymptotic limit
($Q^2 \rightarrow \infty$) the pole mass can be expressed
in terms of the pseudoscalar meson coupling constant
$f_p$ as $f_p^2 = \Lambda_p^2/(8\pi^2)$.
Interpretation of our results in terms of $\Lambda_p$ and $f_p$ and a
comparison
with the previous measurements are shown
in Tables \ref{tab:massresults} and \ref{tab:fpresults}.
Note that the errors quoted are only statistical.
The major source of the systematic error in the pole mass (a relative error
of 5\%) is the experimental uncertainty on $\Gamma_{\gamma\gamma}$.
Other sources of the systematic error include trigger and detector simulation,
tracking, variation in the event selection criteria, background estimates
and an uncertainty in the $Q^2$ of the photon emitted by the missing
electron (positron). Taken in quadrature they correspond to a relative
systematic error of the order of less than 5\% for one-prong and of less
than 10\% for three-prong events.
More careful treatment of the systematic errors is forthcoming.
\begin{table}
\begin{center}
\tcaption{Summary of the results on the $f_p$ [MeV], pseudoscalar meson
coupling constant}
\smallskip
\smallskip
\label{tab:fpresults}
\begin{tabular}{|c|c|c|c|c|}
\hline
& ~~~~Lepton-G\cite{LEPTONG:ff}~~~~ & ~~~~~~TPC/2$\gamma$\cite{TPC:ff}~~~~~
& ~~~~~CELLO\cite{CELLO:ff}~~~~~~ & CLEO (this study) \\
\hline
$\pi^0$ 	  &	         & 	        & 84 $\pm$ 3 & 87 $\pm$ 2 \\
\hline
$\eta$  	  &  81 $\pm$ 10 & 91 $\pm$ 6 & 94 $\pm$ 7 & 89 $\pm$ 3 \\
\hline
$\eta^{\prime}$ &                & 78 $\pm$ 5 & 89 $\pm$ 5 & 92 $\pm$ 2 \\
\hline
\end{tabular}
\end{center}
\end{table}

\section{Summary, Conclusions and Plans}

We reported on the current status of the analysis of
electromagnetic form factors of the light pseudoscalar mesons
$\pi^0$, $\eta$ and $\eta^{\prime}$.
In terms of the pole mass fit and decay coupling constant
our results are consistent with previous measurements.
For the first time $\pi^0$ and $\eta$ form factors
are measured at $Q^2$ above 2.7 and 3.4 ${\rm (GeV/c)}^2$ respectively.
Also, this is the first statistically significant mesurement
for the $\eta^{\prime}$ at $Q^2$ above 8 $(GeV/c)^2$.
With a better understanding of the trigger
we plan to extend our analysis to events with tags
scattering at a smaller angles down to 16.5 degrees.

\end{document}